\documentclass[11pt]{article}

\usepackage[utf8]{inputenc}
\usepackage[T1]{fontenc}
\usepackage{lmodern}
\usepackage[margin=1in]{geometry}
\usepackage{amsmath,amssymb,amsthm}
\usepackage{graphicx}
\usepackage{booktabs}
\usepackage{hyperref}
\usepackage{xcolor}
\usepackage{natbib}
\usepackage{float}
\usepackage{enumitem}
\usepackage{url}
\usepackage{microtype}

\hypersetup{colorlinks=true, linkcolor=blue!70!black, citecolor=blue!70!black, urlcolor=blue!70!black}

\newtheorem{theorem}{Theorem}
\newcommand{\acsinf}{\mathrm{ACS}_\infty}

\title{When Does \texorpdfstring{$q$}{q}-error Predict Plan Regret?\\
Three Regimes of Cardinality-Estimation Error}

\author{
  Madhulatha Mandarapu\thanks{madhulatha@samyama.ai} \and
  Sandeep Kunkunuru\thanks{sandeep@samyama.ai}
}
\date{%
  VaidhyaMegha Private Limited, India\\[2pt]
  \url{https://samyama.ai/}\\[8pt]
  June 2026
}

\begin{document}
\maketitle

\begin{abstract}
Cardinality-estimation (CE) research ranks estimators by \emph{$q$-error}, yet it is well known
that $q$-error is an imperfect proxy for query-plan quality. We give a measurement-driven account of
\emph{when} it is a good proxy and when it is not, and why. Modeling plan selection as an argmin over a
piecewise-linear cost landscape, we find that plan \emph{regret} (the cost of the chosen plan relative to
the optimal, under true cardinalities) is governed by plan-cost geometry in a regime-dependent way.
(i)~For \emph{small} errors, a true-point \emph{condition number} $\kappa$ predicts regret and
out-predicts $q$-error; its predictive power decays to zero as error grows, as a local linearization must.
(ii)~For \emph{large} errors---where deployed learned estimators operate---an estimator-independent
\emph{average-case sub-optimality} measure $\acsinf$ predicts which queries are regret-prone (Spearman
$\rho \approx 0.54$ on STATS-CEB), while $q$-error is nearly uninformative at the query level
($\rho \approx 0.05$). (iii)~The worst case is Haritsa's maximum sub-optimality (MSO). The three are one
cost-ratio spectrum under three weightings. We prove a limit law $\acsinf(q)=\sum_k r_k\pi_k$ with
cardinality-independent combinatorial weights, and validate every claim on STATS-CEB and JOB-light with
four released estimators under pre-registered decision rules, and confirm on \emph{real PostgreSQL runtime}
that $\acsinf$ predicts regret where $q$-error does not. The contribution is conceptual and
empirical---an average-case companion to worst-case robust query optimization, and a characterization of
when an accuracy metric tracks plan quality---rather than a new estimator. Code and the full
pre-registration are public.\footnote{\url{https://github.com/samyama-ai/ce-metric-eval}}
\end{abstract}

\section{Introduction}\label{sec:intro}

A cardinality estimator predicts the size of a (sub-)query result; the query optimizer feeds these
estimates to a cost model and chooses a plan. Errors propagate, and bad estimates yield bad plans
\citep{leis2015job}. The field measures estimator quality with \emph{$q$-error}, $\max(\hat c/c,\,c/\hat c)$,
which enjoys a worst-case guarantee: bounding the maximum $q$-error over a plan's sub-expressions bounds the
plan's sub-optimality \citep{moerkotte2009qerror}. Yet that bound is loose---``impractically large when the
estimation error is significant, as is often the case'' \citep{haritsa2020robust}---and empirical studies
repeatedly find that large $q$-error improvements need not improve plans \citep{han2022statsceb,lee2023impact}.

We ask a sharper question: \emph{when} does a smaller $q$-error mean a better plan, and what \emph{does}
predict plan regret when $q$-error does not? Treating plan choice as an argmin over a piecewise-linear
(in log-cardinality) cost landscape, we find a clean, regime-dependent answer
(Figure~\ref{fig:flow}):

\begin{figure}[t]
  \centering
  \includegraphics[width=\linewidth]{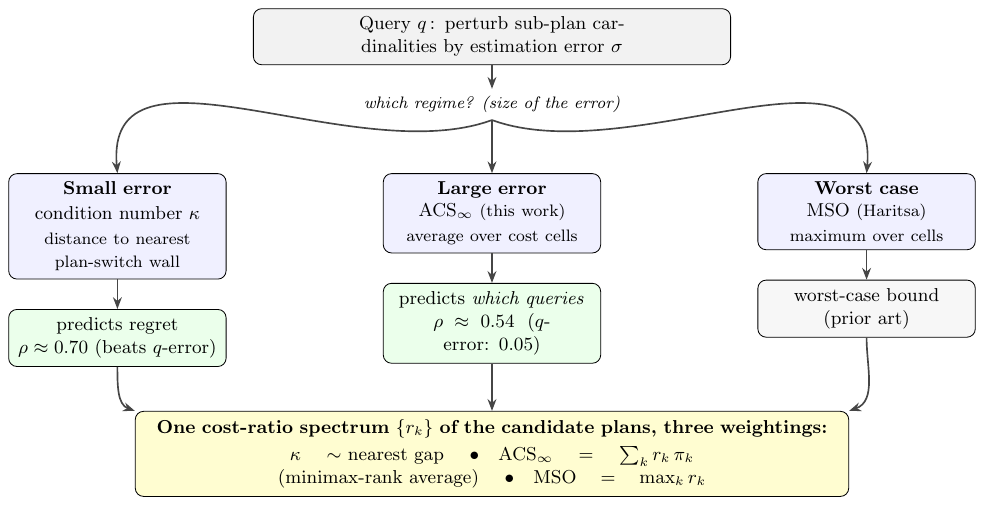}
  \caption{\textbf{The three regimes.} Which plan-cost-geometry quantity predicts plan regret depends on
  the size of the cardinality-estimation error; all three are weightings of the same plan cost-ratio
  spectrum $\{r_k\}$, and $q$-error---an estimate-magnitude scalar---is orthogonal to that geometry.}
  \label{fig:flow}
\end{figure}

\begin{enumerate}[leftmargin=*]
  \item \textbf{Small-error regime.} A per-query \emph{condition number} $\kappa$---the distance to the
  nearest plan-switch boundary---predicts regret and out-predicts $q$-error; its power decays to zero as
  error grows, because the underlying forward$\,\approx\,$condition-number$\,\times\,$backward-error
  relation is local (Section~\ref{sec:small}).
  \item \textbf{Large-error regime.} Where deployed learned estimators sit, local quantities stop working.
  An estimator-independent \emph{average-case sub-optimality} $\acsinf$ predicts which queries are
  regret-prone ($\rho\approx0.54$ on STATS-CEB), against $\rho\approx0.05$ for $q$-error at the query level
  (Section~\ref{sec:large}).
  \item \textbf{A limit law.} $\acsinf(q)=\sum_k r_k\pi_k$, where $r_k$ are true plan cost-ratios and
  $\pi_k$ are cardinality-independent combinatorial weights; the worst-case companion is
  $\mathrm{MSO}(q)=\max_k r_k$ \citep{haritsa2020robust} (Section~\ref{sec:theory}).
\end{enumerate}

This reconciles the long-running ``$q$-error vs.\ plan-cost'' tension as two regimes of one phenomenon.
The result is modest and stands entirely on prior work (Section~\ref{sec:related}); we make no claim to a
new estimator. All experiments use pre-registered decision rules---including one hypothesis we
\emph{rejected} and one near-miss we kept on record---and all code is public.

\section{Setup}\label{sec:setup}

A query $q$ has valid plans $P_1,\dots,P_K$; plan $k$'s internal (join) nodes form a set $I_k$ of
connected table-subsets $S$. Under a cardinality vector $c$ (one entry per subset), the cost is the
$C_{\text{out}}$ model $C_k(c)=\sum_{S\in I_k} c_S$. The optimizer chooses $\hat k=\arg\min_k C_k(\hat c)$
under estimates $\hat c$; the truth-optimal plan is $k^\ast=\arg\min_k C_k(c)$. \emph{Plan regret}
(equivalently P-error / Plan-Cost \citep{negi2021flowloss,han2022statsceb}) is
\[
  \rho(\hat c)=\frac{C_{\hat k}(c)}{C_{k^\ast}(c)}\ge 1,
\]
i.e.\ the chosen plan's cost \emph{under truth} relative to optimal. Let $r_k=C_k(c)/C_{k^\ast}(c)\ge1$ be
plan $k$'s true cost-ratio; $\{r_k\}$ is the query's cost-ratio spectrum. We model estimation error as
$\log\hat c_S=\log c_S+\varepsilon_S$, $\varepsilon_S\sim\mathcal N(0,\sigma^2)$ i.i.d., sweeping $\sigma$
from small (accurate) to large (inaccurate); $q$-error of a subset is $e^{|\varepsilon_S|}$.

\begin{figure}[t]
  \centering
  \includegraphics[width=0.62\linewidth]{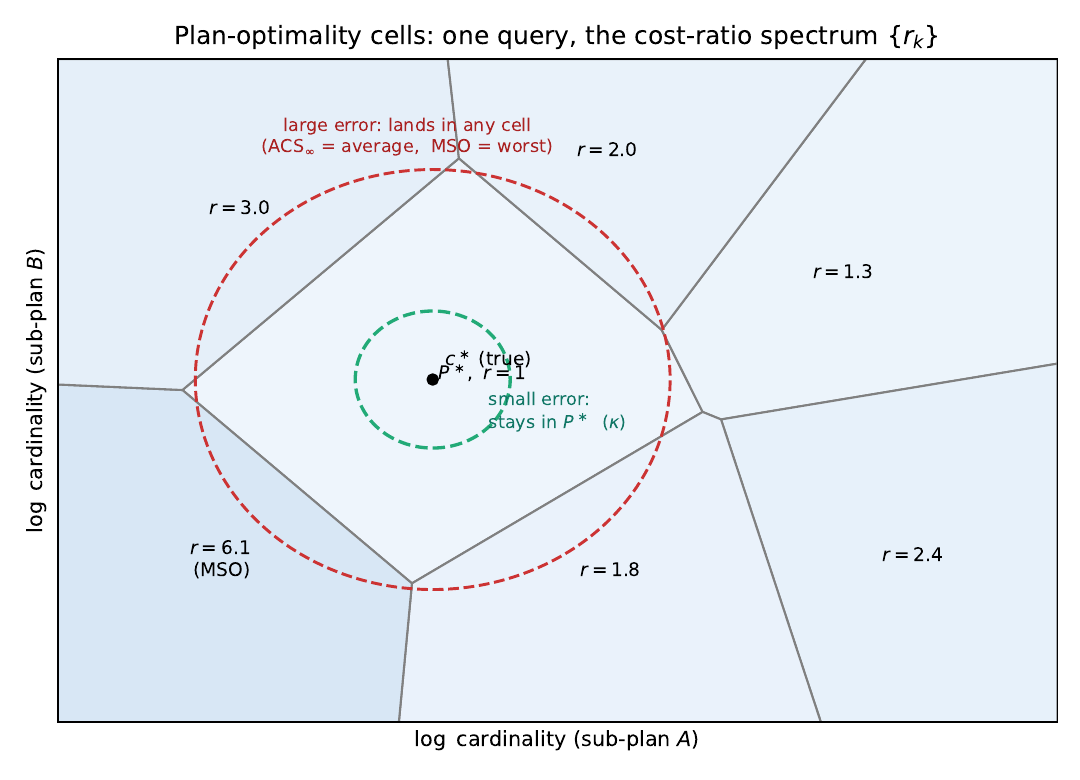}
  \caption{\textbf{Plan-optimality cells.} Log-cardinality space tiles into cells where each plan is
  optimal---a plan diagram \citep{reddy2005plandiagrams}. Small error keeps the estimate in the optimal
  cell, so the condition number $\kappa$ is the distance to the nearest wall; large error lands it in any
  cell, where $\acsinf$ averages the cost-ratios and MSO takes the worst.}
  \label{fig:cells}
\end{figure}

\begin{figure}[t]
  \centering
  \includegraphics[width=\linewidth]{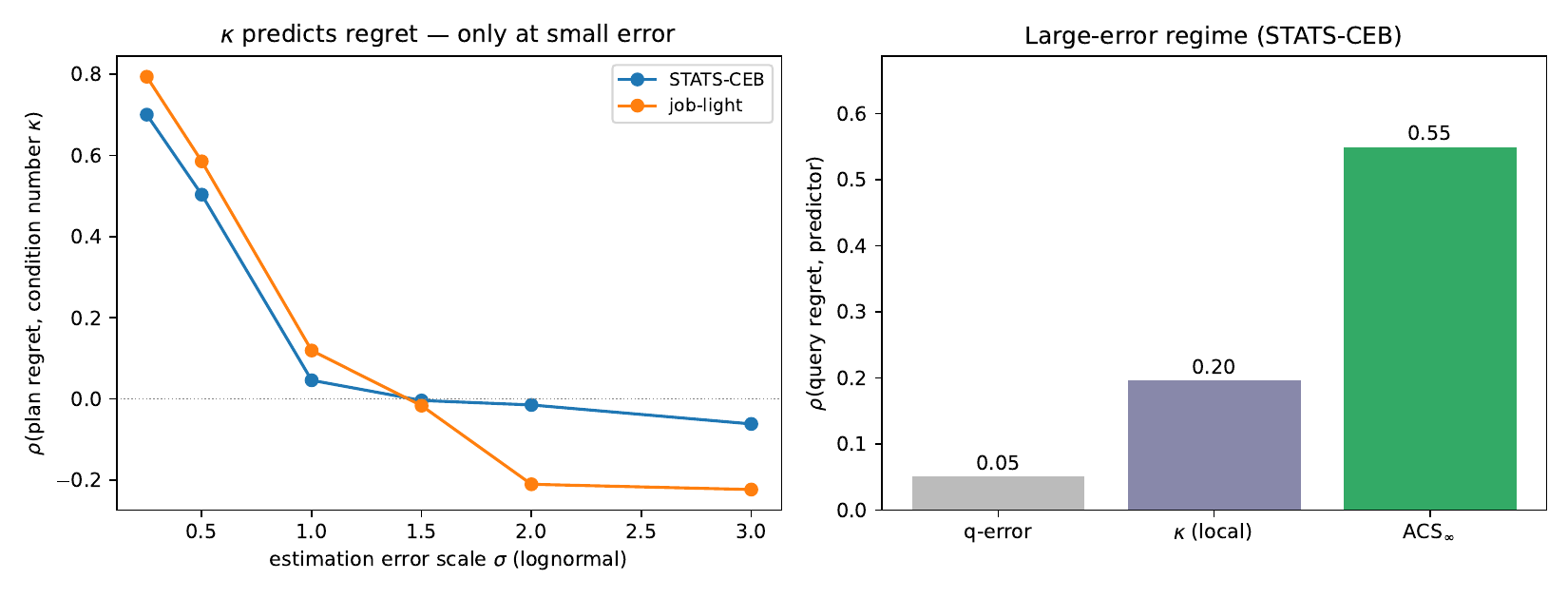}
  \caption{\textbf{Left:} the true-point condition number $\kappa$ predicts plan regret only for small
  estimation error, decaying to $\approx 0$ as error grows (both benchmarks). \textbf{Right:} in the
  large-error regime where deployed estimators operate, $\acsinf$ predicts which queries suffer regret far
  better than $q$-error or $\kappa$ (STATS-CEB, query-level Spearman).}
  \label{fig:regimes}
\end{figure}

\section{Small error: a condition number}\label{sec:small}

\begin{figure}[t]
  \centering
  \includegraphics[width=0.60\linewidth]{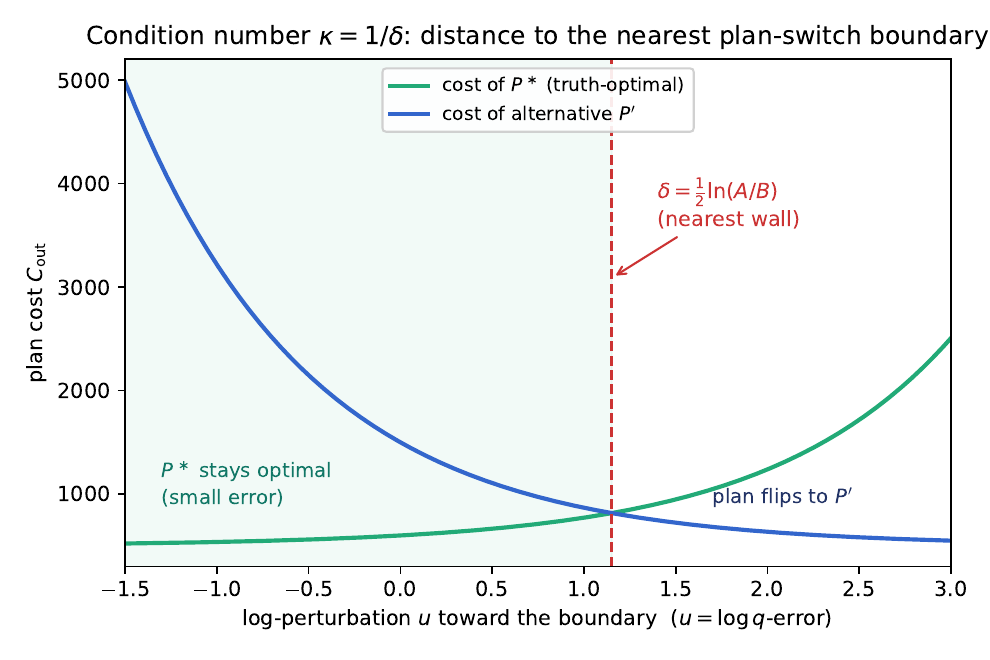}
  \caption{\textbf{The condition number.} Two plans' costs cross at $\delta=\tfrac12\ln(A/B)$; within
  $\delta$ the truth-optimal plan survives, beyond it the choice flips. We set $\kappa=1/\delta$ for the
  nearest such boundary.}
  \label{fig:flip}
\end{figure}

Regret is a forward error; $q$-error is a backward error; the missing factor is a \emph{condition number}.
For the optimal plan $P^\ast$ and an alternative $P'$, the smallest $L_\infty$ log-perturbation that flips
the choice has the closed form $\delta=\tfrac12\ln(A/B)$, where $A,B$ are the true-cardinality sums over
the plans' non-shared internal nodes; we set $\kappa=1/\min_{P'}\delta$. Intuitively, a query whose true
cardinalities sit near a plan-switch boundary is ill-conditioned (Figure~\ref{fig:flip}).

\paragraph{Result.} At small $\sigma$, the per-query Spearman correlation $\rho(\text{regret},\kappa)$ is
$0.70$ on STATS-CEB and $0.79$ on JOB-light (an independent IMDB-schema workload), exceeding the
correlation with realized $q$-error, robust to query size (partial $\rho\approx0.65$ controlling for the
number of tables), and \emph{decaying to $\approx0$ as $\sigma$ grows} (Figure~\ref{fig:regimes}, left).
This is the local prediction made explicit: the condition number governs regret precisely while the
linearization holds. The decay---and the failure of $\kappa$ at large error---is not a defect but a
signature of locality, and it motivates Section~\ref{sec:large}.

\section{Large error: average-case sub-optimality}\label{sec:large}

At large $\sigma$ the estimate is displaced far across the cost-cell complex, so true-point quantities
(\,$\kappa$, and---we verify---cost-weighted and discriminative variants\,) stop predicting regret. We
instead define the \emph{average-case sub-optimality}
\[
  \acsinf(q)=\lim_{\sigma\to\infty}\;\mathbb E_{\varepsilon}\big[\rho(\hat c)\big],
\]
the average-case analogue of Haritsa's worst-case MSO \citep{haritsa2020robust}. Intuitively $\acsinf$ is
``the typical badness of a random plan for this query'': a query whose cost-ratio spectrum is dispersed is
intrinsically regret-prone under poor estimation, \emph{independent of the estimator}.

\paragraph{Result (pre-registered).} On STATS-CEB---where the four released estimators (BayesCard, DeepDB,
FLAT, NeuroCard) are inaccurate (median $q$-error $2$ to ${\sim}10^5$)---$\acsinf$ predicts per-query
regret at $\rho\approx0.54$, against $\rho\approx0.05$ for $q$-error and $\rho\approx0.20$ for $\kappa$
(Table~\ref{tab:large}, Figure~\ref{fig:regimes} right). Under a pre-registered, margin-primary rule, the
bootstrap $95\%$ confidence interval of the margin $\rho_{\acsinf}-\rho_{q}$ is $[0.22,\,0.74]$; the effect
holds on held-out query halves and on a fifth, unseen estimator (the DuckDB optimizer's native estimates);
and it is regime-specific---$\acsinf$ does \emph{not} win on the small-error JOB-light workload, where
$\kappa$ does. Because $\acsinf$ uses no estimator outputs, every test on real estimators is intrinsically
out-of-sample. $\acsinf$ is a \emph{query-level} measure: it predicts which queries CE error endangers,
not estimator-to-estimator variation within a query.

\begin{table}[t]
  \centering
  \caption{Predicting query-level regret in the large-error regime (STATS-CEB, Spearman $\rho$).}
  \label{tab:large}
  \begin{tabular}{lc}
    \toprule
    Predictor & $\rho(\text{regret},\cdot)$ \\
    \midrule
    $q$-error (estimate magnitude)      & $0.05$ \\
    $\kappa$ (true-point, local)        & $0.20$ \\
    $\acsinf$ (average-case, this work) & $\mathbf{0.54}$ \\
    \bottomrule
  \end{tabular}
\end{table}

\begin{figure}[t]
  \centering
  \includegraphics[width=0.50\linewidth]{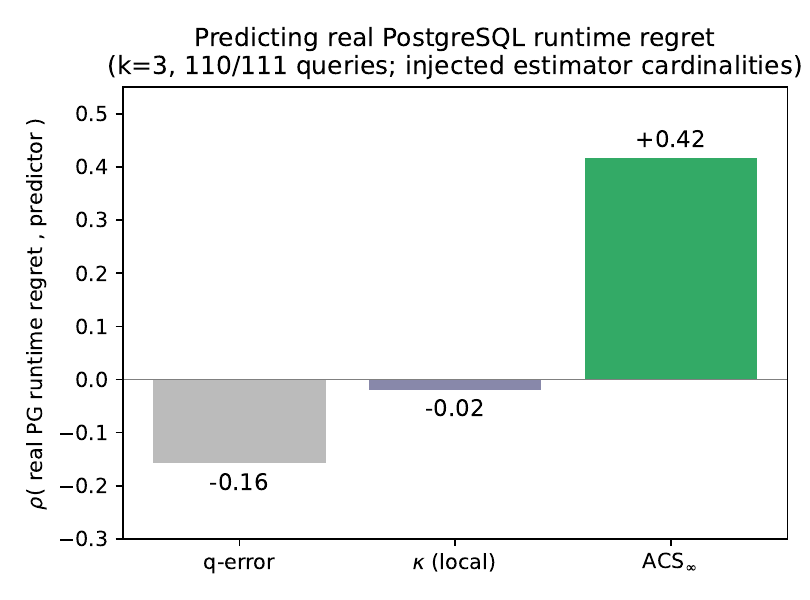}
  \caption{\textbf{Real PostgreSQL runtime.} $\acsinf$ (from $C_{out}$ geometry) predicts actual PostgreSQL
  runtime regret; $q$-error and the local $\kappa$ do not (k=3, 110/111 queries, injected estimator
  cardinalities).}
  \label{fig:runtime}
\end{figure}

\paragraph{Validation on a real optimizer (PostgreSQL).} A natural objection is that $C_{out}$ is a
simplified cost model. We therefore re-measured regret as \emph{actual PostgreSQL~13.1 runtime}: injecting
each estimator's join cardinalities into the optimizer, executing the chosen plan, and taking the ratio to
the true-cardinality plan's runtime (median of three warm runs; a plan exceeding the timeout is recorded as
a regret lower bound, so the highest-regret queries are kept). On full coverage (110/111 queries), $\acsinf$
--- computed purely from $C_{out}$ geometry, never seeing PostgreSQL --- predicts real runtime regret with
$\rho = 0.42$, versus $\rho = -0.16$ for $q$-error, with a bootstrap $95\%$ CI of the margin $=[0.34, 0.82]$
(Figure~\ref{fig:runtime}). The headline regrets are genuine plan changes (e.g.\ $32.6$\,s vs.\ $1.3$\,s).
So the regimes are \emph{not} an artifact of the abstract cost model. (A plan-cost arm via plan pinning is
infeasible: the estimator-induced bad plans are near-cartesian join orders that hint tools will not
reproduce under true cardinalities --- reported as a limitation, not hidden.)

\section{A limit law}\label{sec:theory}

\begin{theorem}[informal]\label{thm:limit}
As $\sigma\to\infty$, optimizer selection converges to the cardinality-free minimax-rank rule
$\hat k\to\arg\min_k\max_{S\in I_k}\varepsilon_S$, and
\[
  \acsinf(q)=\sum_{k=1}^{K} r_k\,\pi_k,
\]
where $r_k$ are the true cost-ratios and $\pi_k$ are \emph{cardinality-independent} selection probabilities
determined solely by the plan set-system $\{I_k\}$.
\end{theorem}

\paragraph{Sketch.} In log-space $\log C_k(\hat c)=\operatorname{LSE}_{S\in I_k}(\log c_S+\varepsilon_S)$,
a log-sum-exp. As $\sigma\to\infty$ the $\varepsilon$-spread dominates the $O(1)$ cost terms, so
$\log C_k(\hat c)\to\max_{S\in I_k}\varepsilon_S$ with probability $\to1$; the argmin then depends only on
the \emph{ranks} of $\{\varepsilon_S\}$ and on which subsets lie in which plans. Hence $\pi_k$ is
combinatorial, and $\acsinf$ \emph{factorizes} into a cardinality part (the cost-ratio spectrum $\{r_k\}$)
and a structure part ($\{\pi_k\}$). The worst-case dual is $\mathrm{MSO}(q)=\max_k r_k$: the same spectrum
under a max instead of the minimax-rank average.

\paragraph{Validation.} The cardinality-aware large-$\sigma$ estimate of $\acsinf$ matches Monte-Carlo at
Spearman $0.99$. Cardinality-independence of $\pi_k$ is confirmed numerically: the correlation between the
cardinality-aware and cardinality-free estimates rises monotonically with $\sigma$ ($0.985$ at $\sigma{=}8$,
$0.990$ at $15$, $0.994$ at $30$), so the residual is finite-$\sigma$ slack, not cardinality dependence.

\section{Related work and reconciliation}\label{sec:related}

This work builds on, and does not originate, the geometry it uses. The partitioning of
selectivity/cardinality space into plan-optimality regions is the \emph{plan diagram} / POSP line
\citep{reddy2005plandiagrams,harish2008robustplans}; our ``cost-cells'' are theirs. The $q$-error bound and
its looseness for large error are due to \citet{moerkotte2009qerror} and restated by
\citet{haritsa2020robust}. Maximum sub-optimality (MSO) and worst-case robust query processing are
Haritsa's; $\acsinf$ is the average-case companion to MSO. \citet{wolf2018robustness} define per-plan
robustness by integrating a \emph{fixed} plan's cost over a cardinality range, for plan \emph{selection};
$\acsinf$ instead measures expected sub-optimality over the optimizer's plan \emph{choice}, as a per-query
difficulty predictor---related but distinct. P-error / Plan-Cost as the regret metric is from
\citet{negi2021flowloss}; the STATS-CEB benchmark from \citet{han2022statsceb}; and the closest empirical
error-injection study is \citet{lee2023impact}, which measures plan-quality impact but does not define a
condition number, a regime crossover, or a $\kappa$-vs-$q$-error comparison.

Against this backdrop the contributions are narrow and specific: the average-case predictor $\acsinf$, its
limit theorem, the $\kappa$/$\acsinf$/MSO regime taxonomy, and the pre-registered demonstration that
$q$-error fails to predict query-level regret in the large-error regime where $\acsinf$ succeeds. The
``$q$-error vs.\ plan-cost'' debate is thus two regimes of one phenomenon; deployed estimators straddle the
boundary, which is why no single scalar error metric predicts their regret cleanly.

\section{Limitations}\label{sec:limits}

Our geometric quantities are defined under the $C_{\text{out}}$ cost model; we validated that the large-error
regime survives on real PostgreSQL runtime (Section~\ref{sec:large}), but the exact crossover location is
cost-model-dependent, and the noise-free plan-cost arm was not obtainable (hint-pinning cannot reproduce the
estimators' near-cartesian plans). Coverage on STATS-CEB is $123/146$ queries (the rest have a
sub-plan join whose true \texttt{COUNT(*)} exceeds a timeout at up to $10^{10}$ rows); JOB-light is fully
covered. $\acsinf$ is a query-level measure and does not predict per-estimator variation. The limit law is a
$\sigma\to\infty$ statement, whereas deployed estimators sit at finite $\sigma$, where a cardinality
correction to $\pi_k$ remains. We invite corrections and pointers to prior art we may have missed.

\section{Conclusion}

Whether $q$-error predicts plan regret depends on the error regime, and in each regime a specific
plan-cost-geometry quantity does the predicting: a condition number for small errors, average-case
sub-optimality for large errors, and maximum sub-optimality for the worst case---one cost-ratio spectrum
under three weightings. The average-case piece, $\acsinf$, is an estimator-independent query-difficulty
measure with a clean limit law, complementing the worst-case robust-query-optimization literature. Code and
the complete pre-registration, including a rejected hypothesis and a near-miss, are public.

\paragraph{Code and data availability.} All code, benchmarks, figures, and the complete pre-registration
trail---including the rejected hypothesis and the near-miss---are public at
\url{https://github.com/samyama-ai/ce-metric-eval}.

\bibliographystyle{plainnat}
\bibliography{paper9_cardinality_regret}

\end{document}